\documentclass[fleqn,12pt,twoside]{article}
\usepackage{espcrc1}
\usepackage{epsfig}
\usepackage{graphicx}

\usepackage{graphicx}
\usepackage[figuresright]{rotating}

\newcommand{\AmS}{{\protect\the\textfont2
  A\kern-.1667em\lower.5ex\hbox{M}\kern-.125emS}}

\title{Electromagnetic break-up of nuclei with $A=3 \div 7$}

\author{G. Orlandini\address{Dipartimento di Fisica, Universit\`a di 
  Trento, and Istituto Nazionale di Fisica Nucleare, Gruppo 
  Collegato di Trento, I-38050 Povo (Trento), Italy}
\thanks{The LIT results presented here are due to the 
work done at the University of Trento with Winfried Leidemann, 
the Ph.D. students Sonia Bacca, Mario A. Marchisio and Sofia Quaglioni, 
and with the external collaborators Nir Barnea, Victor D. Efros and 
Edward L. Tomusiak.}}

\begin{document}

\maketitle

\begin{abstract}
This talk contains a short review of some of the progresses made in the 
last three years in the calculations of electromagnetic cross sections
of light nuclei up to A=7. Since many of them have been possible thanks
to the use of the Lorentz Integral Transform (LIT) method, both for 
inclusive and exclusive reactions, I will first make a few remarks
on the method, stressing its essential points and then show results 
for different nuclei. One of the interesting outcomes is e.g.
the appearing 
of typical collective motion features from ab initio six-body 
calculations. 
When a comparison with available experimental data is attempted,
it is rather disappointing to realize that low-energy data are old, 
incomplete and not accurate enough to disantangle interesting effects, 
showing the need of a major experimental program in this direction,
together with more theoretical efforts to implement modern realistic 
forces in continuum calculations of $A\geq 4$ systems.
\end{abstract}

\section{General remarks on the LIT for inclusive and exclusive cases}

In both inclusive and exclusive electromagnetic reactions one has to 
deal with the very difficult problem of the continuum wave functions 
entering the relevant matrix elements. For light systems this is true 
even at low energies since their discrete spectra are very limited. 
The essential idea of the LIT method is to calculate integral transforms
of these matrix elements (or of proper combinations of them) with 
Lorentzian kernels and then invert the transforms 
(see \cite{ELO94,lapiana00}). 
The reason to take this detour is "economical":  it turns out that, 
in order to calculate the Lorentz transforms of these matrix elements, 
continuum solutions of the Schr\"edinger equation are not required. 
Instead one has to find finite norm solutions to Schr\"odinger-like 
equations with external sources. This implies that one can use 
the much simpler bound state methods to solve them.

For the inclusive case one can prove that one only needs to solve 
\begin{equation}\label{psi1}
  (\hat{H}-E_0-\sigma_R+i\sigma_I )|\widetilde{\psi}_1\rangle = \hat
   {O} | \psi_0\rangle \,,
\end{equation}
while for the exclusive case the solution of the following additional 
equation is required
\begin{equation} \label{psi22}
  (\hat{H}-\sigma_R+i\sigma_I )|\widetilde{\psi}_2\rangle =\hat{V}|
  \phi\rangle 
  \mbox{.}
\end{equation}
In the first equation $E_0$ and $|\psi_0\rangle$ denote the nuclear 
ground state energy and wave function, respectively, $H$ is the 
nuclear Hamiltonian, $\hat O$ is the electromagnetic operator and 
$\sigma_R$ and $\sigma_I$ are the Lorentz kernel parameters, i.e. 
center and width of the Lorentzian, respectively. In the second
equation $\hat V$ denotes the potential between particles belonging to
different fragments and $|\phi\rangle$ is the wave function 
of non interacting fragments. 

In the inclusive case the norm of $|{\psi}_1\rangle$ is the LIT
of the inclusive response function, in the exclusive
case the overlap between $|{\psi}_1\rangle$ and $|{\psi}_2\rangle$ is
connected in a simple way to the matrix element of an exclusive reaction.
I will not report here the derivation of the LIT method which can be found 
e.g. in   \cite{ELO94,lapiana00}. I will only stress that in both 
cases an essential point is the use of the closure property of the 
Hamiltonian eigenstates. This means that, in a way, the LIT method is a 
very powerful extension of sum rule approaches. 


\begin{figure}[htb]
\begin{minipage}[t]{0.47\textwidth}
\includegraphics[width=\textwidth]{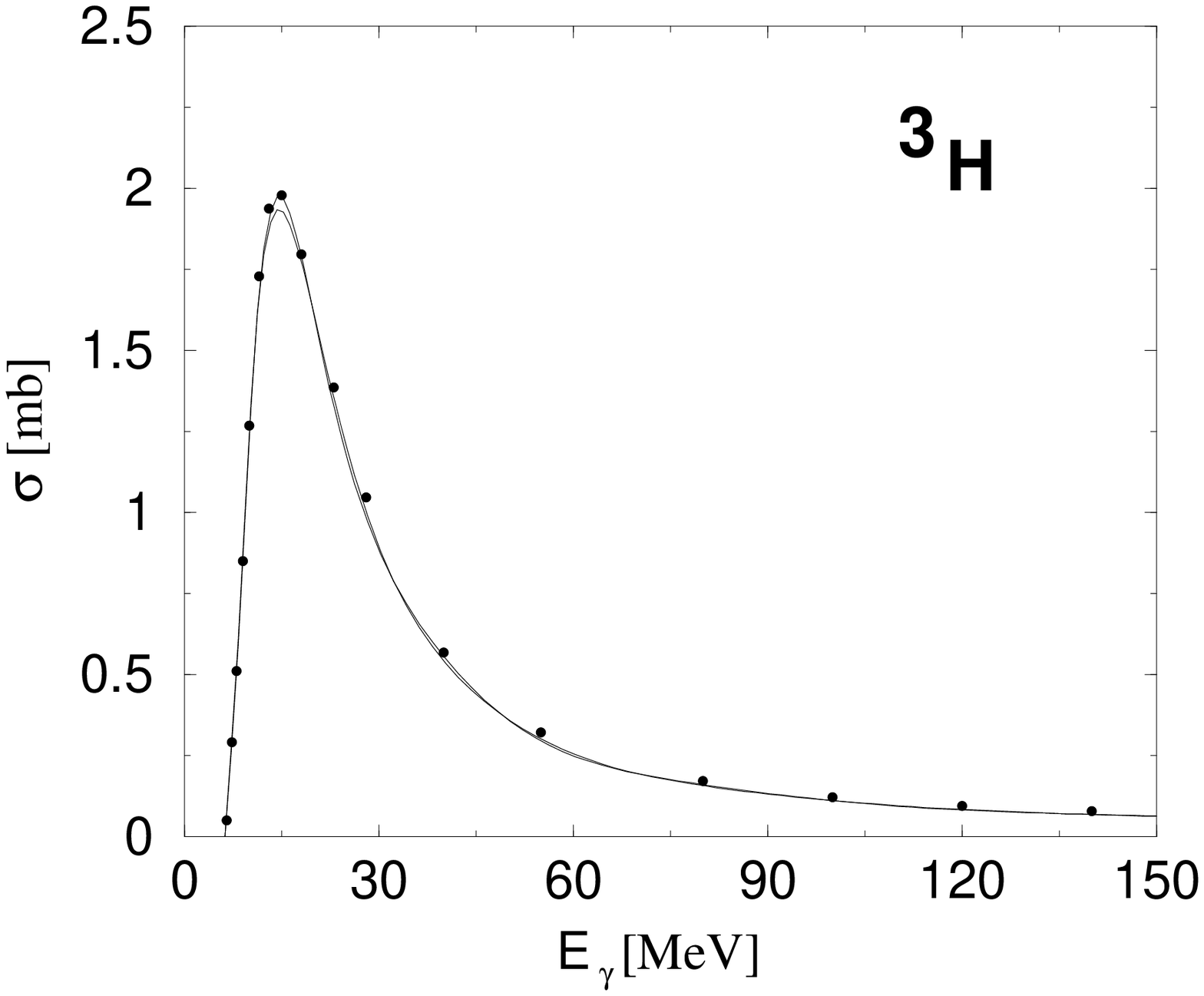}
\caption{Faddeev (dots) and LIT result (upper and lower bounds, 
full curves) in unretarded E1 approximation and for AV18 potential. 
From   \cite{bench02}.}
\end{minipage}
\hspace{\fill}
\vspace{-5mm}
\begin{minipage}[t]{0.47\textwidth}
\includegraphics[width=\textwidth,angle=0]{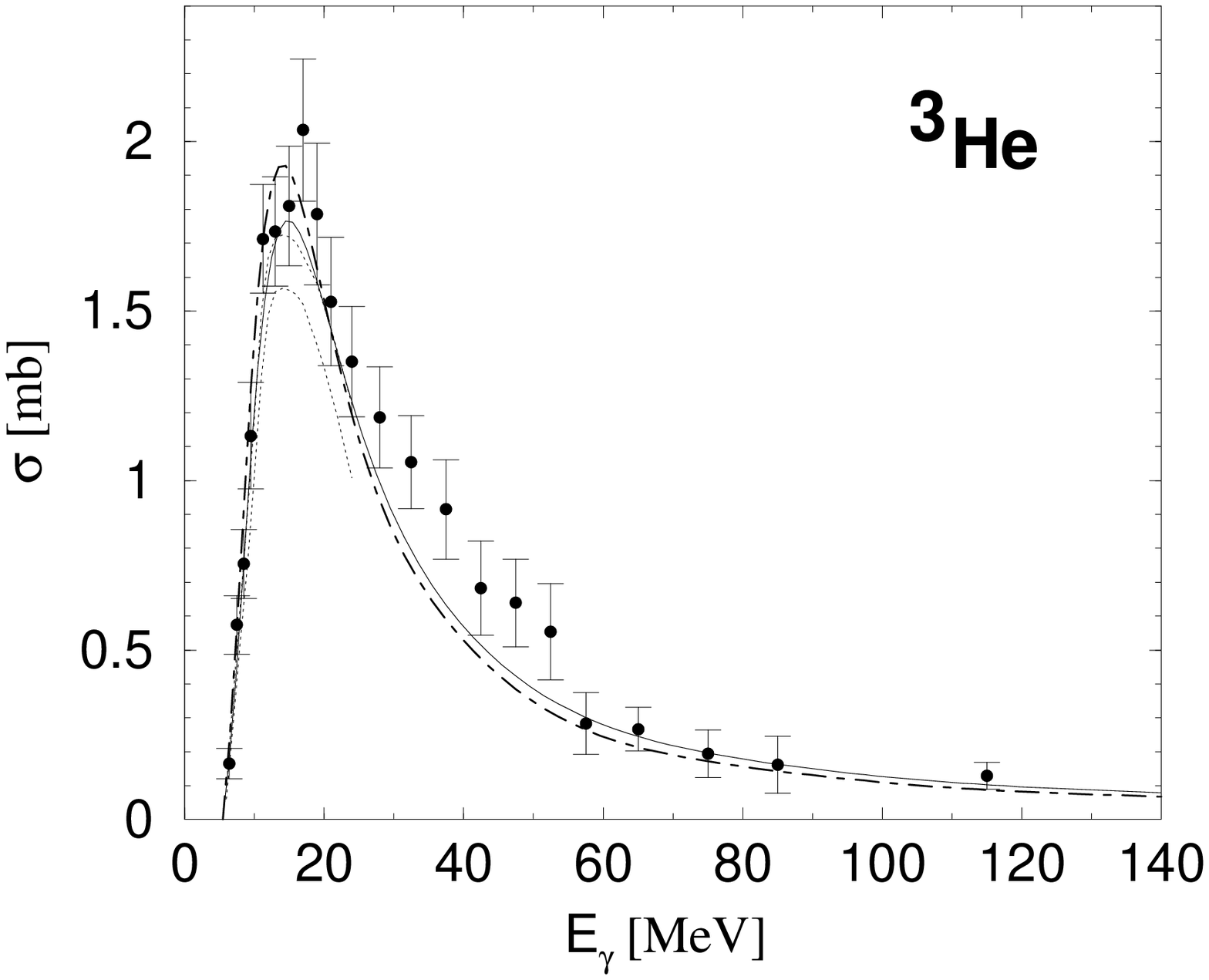}
\caption{Total $^3$He photoabsorption cross section. LIT results in 
unretarded E1 approximation with AV18 only (dash-dotted curve) and 
with AV18+UIX (full curve) compared with two sets of data 
(see \cite{bench02}).}
\end{minipage}
\end{figure}


It may seem that the difficulties of finding solutions in
the continuum are translated into the difficulties of inverting integral
transforms. However, it turns out that for the case of a Lorentzian 
kernel the inversion does not suffer from the uncontrolled instabilities 
typical of other kernels (like e.g. the Laplace one) and 
that it is possible to obtain very accurate results. This can be seen
comparing LIT results  with those obtained using explicit continuum 
states. The comparison has been possible only for cases where explicit 
continuum states were available, i.e. the two- and three-body systems 
(see \cite{ELO94,bench02}).
As an example in Fig.~1 the results obtained in  \cite{bench02} 
are reported,
where the photonuclear cross section of triton has been calculated
in unretarded dipole approximation, both by solving the Faddeev equations 
and by the LIT method. As one can see the agreement is at the level of 
few percents. In Fig.~2 the results are compared with available
experimental data. It is clear that more accurate experiments are needed
if one wants to disantangle the effects of three-body forces.

In the following I will review results obtained in larger systems, where
the LIT seems to be at present the only method to calculate
electromagnetic cross sections in a large energy range, especially
beyond the two-body break-up thresholds.

\section{Results for nuclei with $A>3$}

The photonuclear two-body break-up of $^4$He has been calculated 
(see Fig.~3) applying the exclusive version of the LIT method.  
This allows to give for the first time a result where the final 
state interaction is fully taken into account, also
beyond the three-body break-up threshold. The potential has been 
chosen to be simply central (MTI-III, \cite{MT69}). The experimental 
situation is very complicated. More comments about the comparison 
between theory and experiment can be found in the contribution by 
S. Quaglioni to this conference.


\begin{figure}[htb]
\vspace{-5mm}
\centering\includegraphics[width=0.7\textwidth]{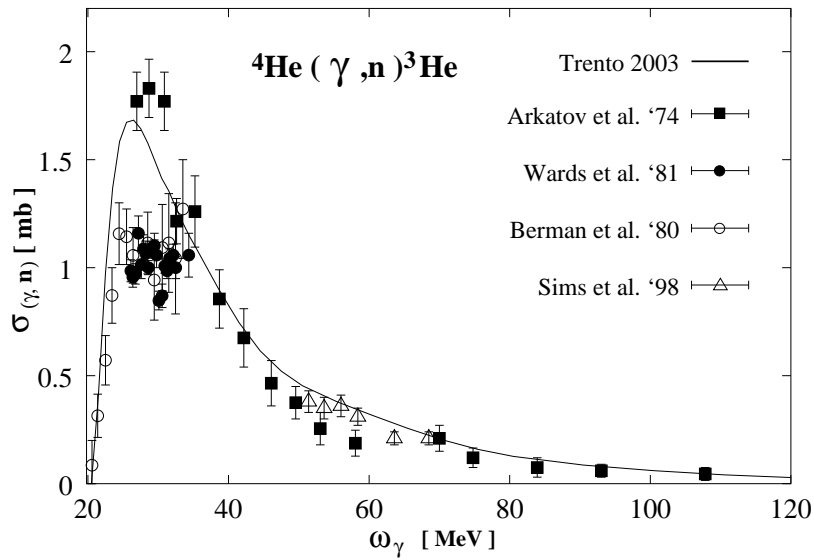}
\caption{Comparison between the theoretical result for the 
$^4$He$(\gamma,n)$ photonuclear cross section and data from 
various experiments (for references see the contribution by 
S. Quaglioni to this conference)}
\end{figure}


\begin{figure}[htb]
\vspace{-10mm}
\includegraphics[width=\textwidth]{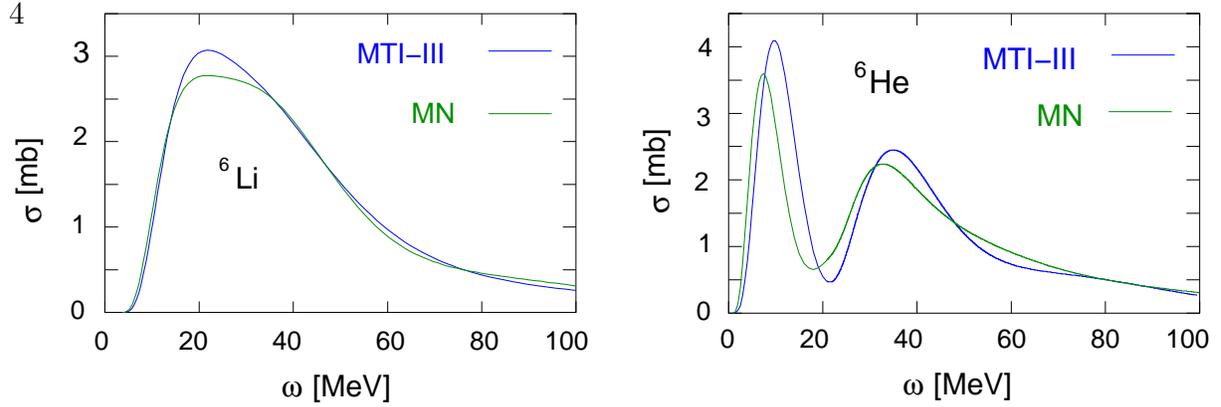}
\vspace{-10mm}
\caption{Total $\gamma$-cross sections of $^{6}$Li and $^{6}$He
with MTI-III and MN \cite{MN77} potentials.}
\end{figure}


\begin{figure}[htb]
\includegraphics[width=\textwidth]{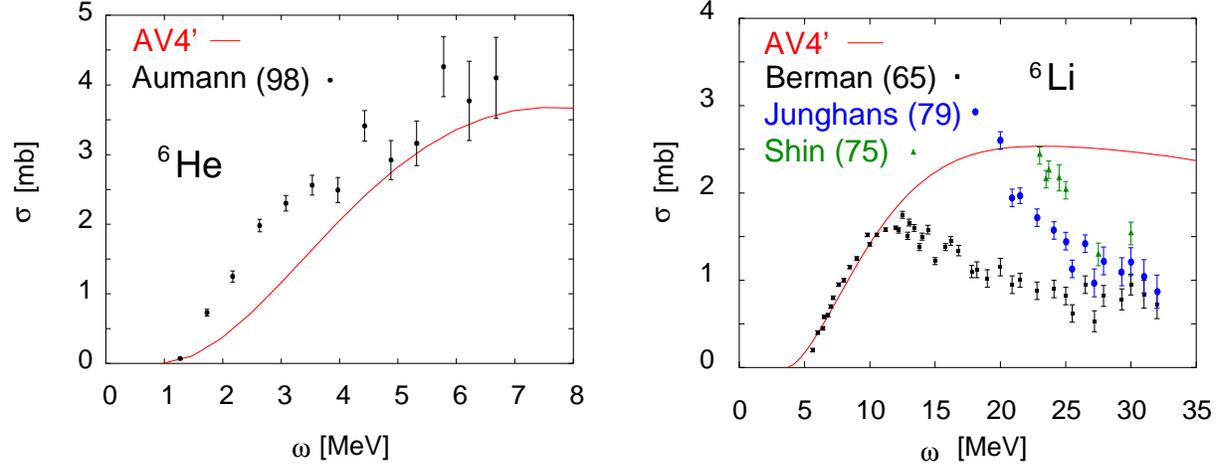}
\caption{Total $\gamma$-cross sections of $^{6}$Li and $^{6}$He with AV4'
potential compared with data (for references see \cite{bacca02}).}
\vspace{-5mm}
\end{figure}


\begin{figure}[htb]
\begin{minipage}[t]{0.47\textwidth}
\includegraphics[width=\textwidth, angle=0]{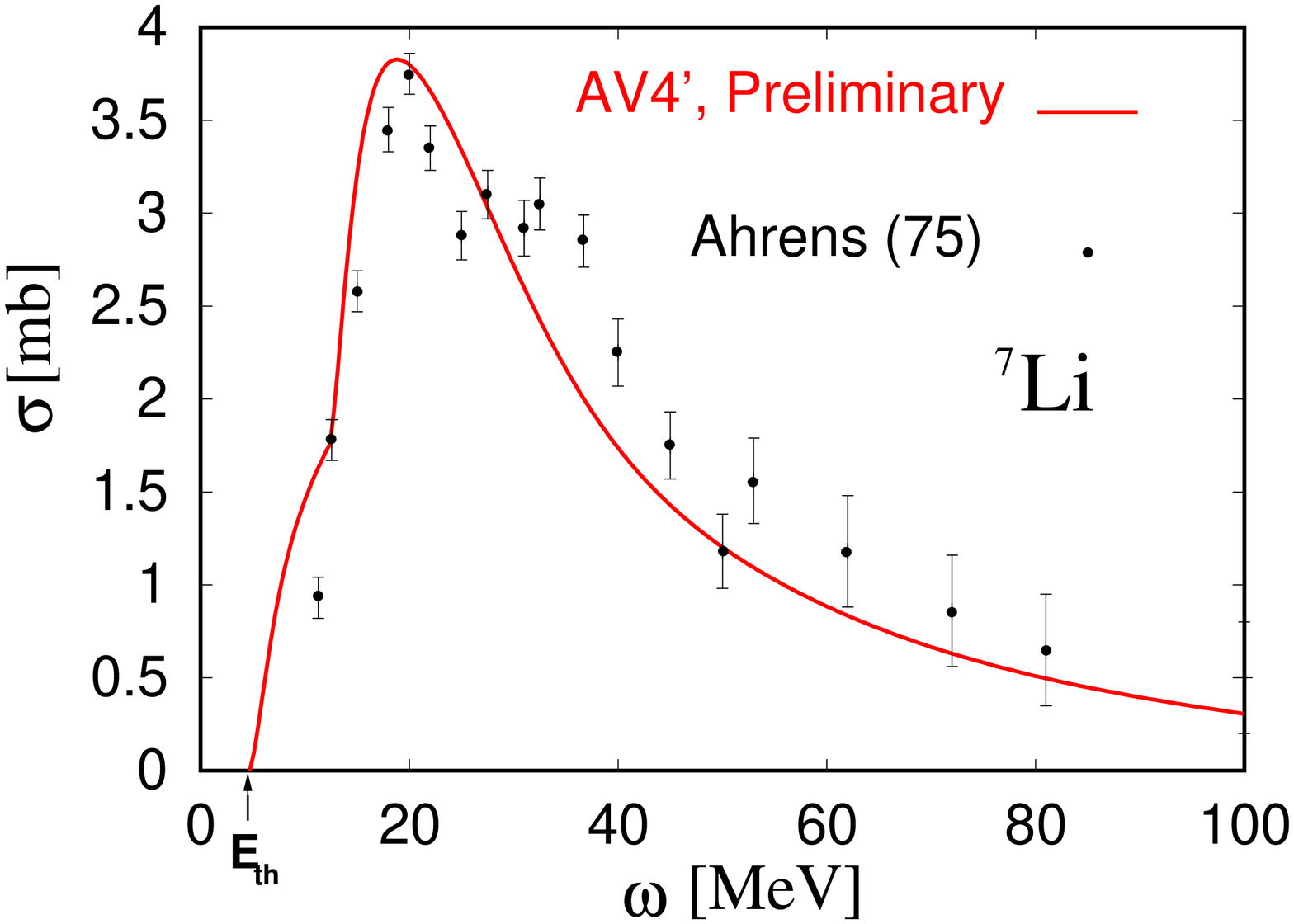}
\caption{Total $\gamma$ cross sections of $^{7}$Li with AV4'
potential compared with data \cite{ahrens75}.}
\label{fig:largenenough}
\end{minipage}
\hspace{\fill}
\vspace{-5mm}
\begin{minipage}[t]{0.47\textwidth}
\includegraphics[width=\textwidth,angle=0]{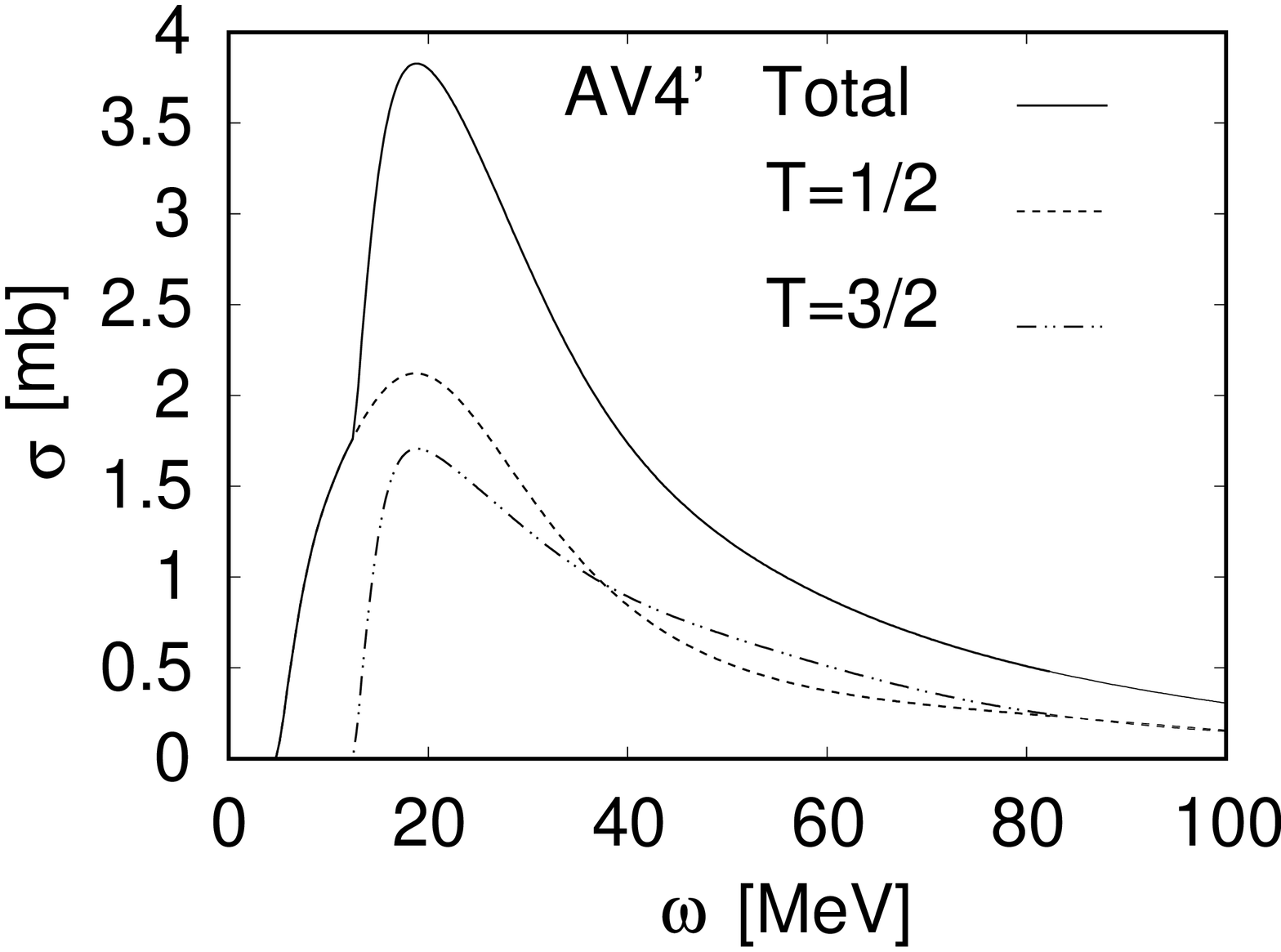}
\caption{T=1/2 and T=3/2 contributions to the total
$\gamma$ cross sections of $^{7}$Li.}
\label{fig:toosmall}
\end{minipage}
\end{figure}

In Fig.~4 inclusive results on the total dipole photodisintegration 
cross sections of $^6$He and $^6$Li are presented (see also   
\cite{bacca02}). A noticeable 
feature is the appearing of two resonance peaks in the $^6$He case.
They may be interpreted as the "soft dipole mode" due to the 
collective oscillation of the neutron halo against the 
alpha core and the classical Gamow-Teller mode of the protons
against the neutrons. Of course in the latter case the break-up of 
the alpha-core is required and therefore this mode appears at
higher energy. 

It is very interesting to see a collective
feature stemming out of a six-body microscopic calculation.
This feature seems to persist independently on the potential
used. In principle these two modes might exist also in the case
of  $^6$Li, (in this case the soft mode could be that of the 
n-p pair against the alpha particle). One possible explanation 
for the unique peak of $^6$Li could be the existence of an
additional dipole mode filling the gap between the soft and the 
Gamow-Teller mode. This third mode would correspond to a 
$^3$He-$^3$H oscillation. The corresponding $^3$H-$^3$H channel 
in $^6$He is absent because it is forbidden in dipole approximation. 

In Fig.~5 a comparison with available data is shown. In this case 
calculations have been performed with the  AV4' potential
\cite{AV4P}, which includes P-wave interactions. 
A good agrement with data is obtained for $^6$Li at lower energy.
The P-wave potential contribution is crucial for 
the agreement (compare with Fig.~3 in  \cite{bacca02}). 
Sizeable disagreement
is still found for $^6$He and $^6$Li at higher energies. 
New and more accurate data and calculations with more realistic 
potentials are needed to draw some conclusions about the role of 
the different potential terms.

In Figs. 6 and 7 preliminary results of a microscopic seven-body 
calculation 
of the total photodisintegration of $^7$Li are presented. Also in this 
case the theoretical result shows a unique peak at about the same energy 
as the experimental data. Also the height of the peak is well reproduced.
The shoulder below 15 MeV is due to different thresholds for the 
T=3/2 and T=1/2 channels, as shown in Fig.~7.

Finally it should be mentioned that Eqs.(1-2) have been solved by 
correlated hyperspherical harmonics (CHH) expansions for A=3,4.
For A=6,7 the EIHH method \cite{EIHH} (using the concept of 
effective interaction) and a reformulation of the LIT, permitting to take 
advantage of the Lanczos algorithm \cite{lanczos}, have helped  
the convergence of the results.

\end{document}